\documentclass[%
 reprint,
 amsmath,amssymb,
 aps,
]{revtex4-2}

\usepackage{graphicx}
\usepackage{dcolumn}
\usepackage{bm}

\begin{document}

\preprint{APS/123-QED}

\title{Reflection, transmission and surface susceptibility tensor in two-dimensional materials}
\author{Luca Dell'Anna$^{1}$}
\author{Yu He$^{1,2}$} 
\author{Michele Merano$^{1,}$}%
\email{michele.merano@unipd.it}
\affiliation{
 $^{1}$ Dipartimento di Fisica e Astronomia G. Galilei, Universit$\grave{a}$ degli studi di Padova, via Marzolo 8, 35131 Padova, Italy\\$^{2}$ College of Physics, Sichuan University, Chengdu 610064, China.
}

\date{\today}

\begin{abstract}
In a recent experiment, the out-of-plane surface susceptibility of a single-layer two-dimensional atom crystal in the visible spectrum has been measured. This susceptibility gives a measurable contribution to the reflectivity of two-dimensional materials. Here we provide a complete theoretical description of the reflective properties, considering incoming $s$ and $p$ polarized plane waves at any angle of incidence on the crystal, computing local, reflected and transmitted electromagnetic fields. We finally connect the microscopic polarizability to both the in-plane and the out-of-plane macroscopic surface susceptibilities.
\end{abstract}

\maketitle

\section{INTRODUCTION}

Two-dimensional (2D) crystals are the thinnest materials that can be produced \cite{2004electric, Novoselov2005}. They are composed of only one atomic plane like graphene or Boron-Nitride \cite{Blake2011, VandeWalle2018}, or one molecular plane like transition metal dichalcogenides \cite{mak2010atomically,li2014measurement}. This planar configuration make them highly anisotropic in the direction perpendicular to the crystal plane. For instance, they can have in-plane macroscopic dimensions \cite{Huang2015, Magda2015, Desai2016, huang2020universal} while keeping an out-of-plane thickness of the order of one atom. This particular geometry suggests that also their optical response should be anisotropic, at least in the vertical direction to the crystal plane. Ab-initio calculations indeed predict this anisotropy for a 2D crystal \cite{trevisanutto2010ab,guilhon2019out, Guo20}. Notwithstanding the enormous progresses in thin films and 2D materials optical characterization \cite{chang2014extracting, Liu14, Li14, stenzel2015physics, tompkins2015spectroscopic,  Morozov15, Jayaswal18, Elliott20}, it turns out that a measurement of the out-of-plane optical constants of a 2D crystal is still a difficult task \cite{nelson2010optical}.  

These materials are usually deposited on a substrate whose optical response is added to that of the monolayer. This is enough to hide the contribution coming from the out-of-plane optical constants of the single-layer crystal \cite{nelson2010optical}. In thin-film optics, the sensitivity of a measurement to vertical anisotropy is dependent on the path length of the light through the film, which is extremely limited when we deal with these atomically thin crystals \cite{nelson2010optical}. As a consequence, optical experiments of monolayers, deposited on some substrate, have access only to the in-plane optical constants of the samples that are studied \cite{li2014measurement, chang2014extracting, Liu14, Li14, Morozov15, blake2007making, Jayaswal18, Elliott20}.  

Only recently an experiment successfully measured the out-of-plane optical constants of 2D crystals, namely graphene and monolayer $\rm MoS_2$ \cite{xu2021optical}. The substrate contribution was eliminated by a complete immersion of these single-layer crystals in a transparent polymer. Experimental data say that the out-of-plane surface susceptibility ($\chi_{\bot}$) is a measurable and finite quantity different from the in-plane surface susceptibility ($\chi _{\Vert}$), while the out-of-plane surface conductivity was zero within the experimental errors.

One might expect that a 2D crystal with atomic thickness along the vertical direction has a vanishing $\chi _{\bot}$, being $\chi _{\bot}$ a macroscopic quantity. Indeed some of the most cited papers about optics in 2D crystals  assume a null $\chi _{\bot}$ \cite{ falkovsky2007optical, hanson2008dyadic, Zhan2013, merano2016fresnel}. All the other papers describe the linear optical response of a monolayer assuming isotropy even in the vertical direction \cite{blake2007making,  nelson2010optical,li2014measurement}.  
The aim of this work is to prove that $\chi _{\bot}$ is indeed finite and different from $\chi _{\Vert}$ in a rigorous analytical way.

We will start with a microscopic description of a 2D crystal, as done in Ref. \cite{dell2016clausius}, and then will connect it to the macroscopic one, by means of a procedure similar to that used in Ref. \cite{merano2017role}. The results in Refs. \cite{dell2016clausius, merano2017role} were limited to normal incidence, therefore, they had no access to $\chi_\bot$.

\section{MICROSCOPIC THEORY}

\begin{figure}
\centerline{\includegraphics[width=8.cm]{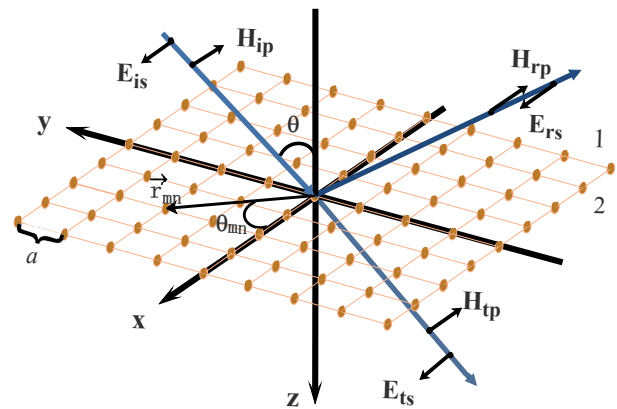}}
\caption{\label{Fig1}  $s$- ($p$-) polarized electric (magnetic) field incident with an angle $\theta$ on a single-layer two-dimensional atomic crystal, modeled by dipoles placed on a regular lattice in the $x$-$y$ plane. The two half-spaces, separated by the crystal, are denoted by the labels $1$ and $2$.}
\end{figure}

We consider an insulating 2D crystal, in the vacuum, composed by atoms with isotropic polarizability $\alpha$ placed on a 2D Bravais lattice (Fig.\ref{Fig1}). A plane wave propagating in a generic direction specified by the unit vector $\hat{s}=\cos\phi\sin\theta\, \hat{x} -\sin\phi\sin{\theta}\, \hat{y} +\cos{\theta} \, \hat{z}$ which is defined by the polar angles $\theta$ (the incident angle) and $\phi$ (the azimuthal angle). Where $\hat{x}$, $\hat{y}$, $\hat{z}$ are unit vectors along the $x$, $y$ and $z$ directions. The incident electric field varies periodically in time $\vec{E}_{i}(t) = \vec{E}_{i}\,e^{i\omega t}$ 
where $\omega$ is the frequency from which one can define $k=\omega / c=2\pi/\lambda$, where $\lambda$ is the wavelength of 
the incident field. The charge distribution in the crystal is distorted by the electric field, which generates oscillations of the electric dipoles placed at the Bravais lattice sites. The total electric field which acts on a single dipole is the local field $\vec{E}_{loc} $. It is the superposition of the incident field $\vec{E}_{i}$ and the contributions arising from all the other dipoles. Because of translation invariance, we make the hypothesis that the local field at any lattice point has the same magnitude and have the same frequency of the incident field but with a different phase.

\subsection{Computation of the local field}
\begin{figure}
\centerline{\includegraphics[width=8cm]{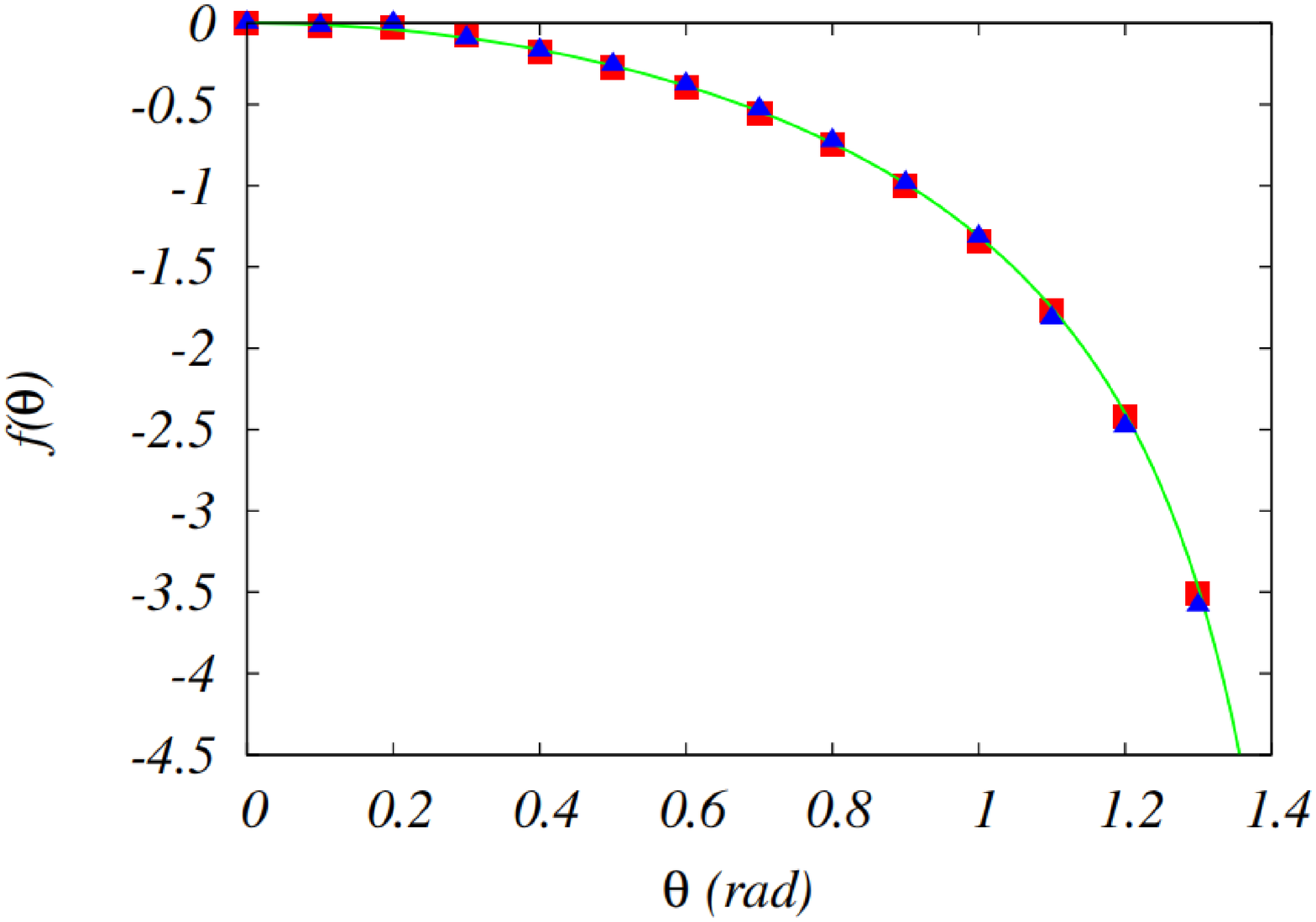}}
\caption{\label{Fig2}$f(\theta)$ as a function of the angle of incidence $ \theta$. The square dots refer to the square lattice  and the triangular dots refer to triangle lattice. The dots are numerical results, summing over almost $10^4$ sites, while the line is given by Eq. (\ref{Ftheta}).}
\end{figure}

We compute the local field acting on the dipole at the origin of our reference system (Fig.1). The contributions to this local field are given by the fields coming from the oscillating dipoles placed at the points labelled by the integers $m$ and $n$, which are given by \cite{Feynmann1964,jackson1999classical,born2013principles}

\begin{align}
\vec{E}_{mn}  = \,&
\frac{{\cal P}_{mn}}{4\pi\epsilon _{0} r_{mn}^{3} } \left[ 3 (\tilde{\vec{p}}\cdot \hat{{r}} _{mn}) \hat{{r}} _{mn}- \tilde{\vec{p}} 
\right.\nonumber \\&\left. 
-\frac{1}{c^{2}} (\hat{{r}} _{mn}\times\ddot{\vec{p}}) \times \hat{{r}} _{mn} \right]  
\end{align}
where $\tilde{\vec{p}}=\vec{p}(t-\frac{r_{mn}}{c})+\frac{r_{mn}}{c}\dot{\vec{p}}(t-\frac{r_{mn}}{c})$.  $\vec{p}$ is the induced dipole moment at each lattice  point, $\epsilon _{0}$ is the vacuum permittivity, $c$ is the velocity of light in vacuum, and 
$\vec{{\emph{r}}}_{mn}=r_{mn}(\cos\vartheta_{mn}\hat{{\emph{x}}}+\sin\vartheta_{mn}\hat{{\emph{y}}})$ while 
$\hat{{\emph{r}}}_{mn}=\vec{{\emph{r}}}_{mn}/r_{mn}$. 
The phase in the prefactor is given by
\begin{equation}
{\cal P}_{mn}=e^{i(\omega t-k\,r_{mn})}\,e^{-ik \delta_{mn}}
\end{equation}
with a phase shift $\delta_{mn}$ whose modulus is the distance of the point $(r_{mn}\cos\vartheta_{mn}, r_{mn}\sin\vartheta_{mn}, 0)$ from 
the plane wave crossing the origin of the reference frame with equation $\hat{s}\cdot \vec{r}=0$.  
As a result, in quite general terms, the phase shift $\delta_{mn}$ can be written as 
\begin{equation}
\delta_{mn}=r_{mn}\sin\theta \cos(\vartheta_{mn}+\phi)
\label{Delta}
\end{equation}
The components of $\vec{{\emph{E}}}_{m n}$ are, therefore,
\begin{widetext}
\begin{eqnarray}
\label{Ex}
&& E_{m n, x}=\frac{e^{i\omega t}}{4\pi \epsilon_{0}\,r_{mn}^3} \,e^{-ik\,(r_{mn}+\delta_{mn})}
\left\{p_x
\left[\left(3\cos^2\vartheta_{mn}-1\right)(1+ik\, r_{mn})+ k^2 r_{mn}^2 \sin^2\vartheta_{mn}\right]\right.\\\nonumber
&& \phantom{------------------}
+\left.p_y\cos\vartheta_{mn}\sin\vartheta_{mn}\left[3(1+k\,r_{mn})-k^2 r_{mn}^2\right]\right\}\\
\label{Ey}
&& E_{m n, y}=\frac{e^{i\omega t}}{4\pi \epsilon_{0}\,r_{mn}^3} \,e^{-ik\,(r_{mn}+\delta_{mn})}
\left\{p_y
\left[\left(3\sin^2\vartheta_{mn}-1\right)(1+ik\, r_{mn})+k^2 r_{mn}^2 \cos^2\vartheta_{mn}\right]\right.\\\nonumber
&& \phantom{------------------}
+\left.p_x\cos\vartheta_{mn}\sin\vartheta_{mn}\left[3(1+k\,r_{mn})-k^2 r_{mn}^2\right]\right\}\\
&& E_{m n, z}=\frac{e^{i\omega t}}{4\pi \epsilon_{0}\,r_{mn}^3} \,e^{-ik\,(r_{mn}+\delta_{mn})}
p_z\left[
k^2 r_{mn}^2 -1-ik\, r_{mn}\right]
\end{eqnarray}
\end{widetext}
Without loss of generality we will take $\phi=\pi/2$ (see Fig. \ref{Fig1}) so that the phase shift reduces to 
\begin{equation}
\delta_{mn}=-r_{mn}\sin\theta \sin\vartheta_{mn}
\end{equation}
Summing over $\vartheta_{mn}$, the last terms in Eqs.~(\ref{Ex}), (\ref{Ey}) are zero, since those terms are odd under $\vartheta_{mn}\rightarrow \vartheta_{mn}+\pi$ for our choice of the reference frame.
As a result the sum over the contributions to the local filed coming from all the dipoles is parallel to the dipole sitting at the origin. \\
For the square lattice $r_{mn}=a\sqrt{m^2+n^2}$, where $a$ is the lattice parameter, and $\tan\vartheta_{mn}=n/m$, while  
for the triangular lattice $r_{mn}=a\sqrt{(m+n/2)^2+3n^2/4}$ and $\tan\vartheta_{mn}=\sqrt{3}\,n/(2m+n)$. After summing over the dipoles, 
and using 
\begin{equation}
\label{pEloc}
\vec{{\emph{p}}}=\alpha\epsilon_0\,\vec{{\emph{E}}}_{loc}
\end{equation}
we get numerical evidences that  
\begin{eqnarray}
\label{Sx}
&&\sum^\prime_{m,n}E_{m n, x}= \frac{\alpha}{4\pi a^3} \left(C_0+i ka \frac{C_1}{\cos\theta}\right)E_{loc, x}\\
\label{Sy}
&&\sum^\prime_{m,n}E_{m n, y}= \frac{\alpha}{4\pi a^3} \Big(C_0+i k a\,C_1\cos\theta\Big)E_{loc, y}\\
\label{Sz}
&&\sum^\prime_{m,n}E_{m n, z}= \frac{\alpha}{4\pi a^3} \Big(\hspace{-0.05cm}-\hspace{-0.05cm}2C_0+i k a \, C_1 f(\theta)\Big)E_{loc, z}
\end{eqnarray}
where the prime on the summation symbol indicates that the origin, $n=m=0$, is excluded from
the sum. $C_{0}$ is the static result already reported in our previous paper \cite{dell2016clausius} ($C_{0} \approx  4.517$ for the square lattice and
$C_{0} \approx  5.517$ for the triangular lattice) and $C_{1} = -2 \pi N a^{2}$, where $N$ is the density of dipoles ($N=1/a^{2}$ for the square lattice, $N=2/\sqrt{3}a^{2}$  for the triangular lattice). \\
We notice that the $\theta$-dependence vanishes in the static limit, namely in the long-wavelength limit, $k\rightarrow 0$, and appears in the three field components with a $1/\cos\theta$, $\cos\theta$ and a combination of them 
\begin{equation}
\label{Ftheta}
 f ( \theta) = \frac{1}{\cos \theta }-\cos \theta = \sin \theta \tan \theta.
\end{equation}
As shown in Fig. \ref{Fig2}, the $\theta$-dependence in Eq. (\ref{Sz}) is the same for either square and triangular lattices. Analogous checks have been performed to verify $1/\cos\theta$ and $\cos\theta$ in Eqs. (\ref{Sx}) and (\ref{Sy}) respectively.
The local field at one site is due to the incident field and the contributions from the other oscillating dipoles. We has, therefore, to solve the following equation 
\begin{align}
\vec{E}_{loc}=\vec{E}_{i}+\sum_{m,n}' \vec{E}_{mn} 
\end{align}
We can now easily find the local field $\vec{E}_{loc}$ from Eqs. (\ref{Sx}), (\ref{Sy}), (\ref{Sz}) getting
\begin{eqnarray}
\label{Elocx}
&&E_{loc,x} = \frac{E_{ix}}{1-\frac{C_{0}\alpha}{4\pi a^{3}}+ik\frac{\alpha N}{2}\frac{1}{\cos \theta }}  \\   
\label{Elocy}
&&E_{loc,y} = \frac{E_{iy}}{1-\frac{C_{0}\alpha}{4\pi a^{3}}+ik\frac{\alpha N}{2}\cos \theta }   \\ 
\label{Elocz}
&&E_{loc,z} = \frac{E_{iz}}{1+\frac{C_{0}\alpha}{2\pi a^{3}}+ik\frac{\alpha N}{2}\sin  \theta\tan \theta }   
\end{eqnarray}
These equations show that the local field, and therefore also the macroscopic polarization vector, as we will see, are connected to the incident field via a diagonal tensor.

\section{MACROSCOPIC THEORY}
In this section, we will find the expression for the reflected and transmitted fields and the relation in between the microscopic polarizability $\alpha$ and the macroscopic surface susceptibility of the crystal. Because of momentum conservation, the reflected field will propagate along  the unit vector $\hat{s}_r=-\sin{\theta}\ \hat{y} -\cos{\theta} \ \hat{z}$, and the transmitted field along the same unit vector $\hat{s}$ of the incident field. 

The first macroscopic quantity that we can introduce is the surface polarization $\vec{P}$ 
\begin{equation}
\label{P}
\vec{P} =N\vec{p} =\epsilon_0 \alpha N\vec{E}_{loc} 
\end{equation}
using Eq.~(\ref{pEloc}) and it is related to the incident field from Eqs.~(\ref{Elocx})-(\ref{Elocz}).  
The surface susceptibility is a diagonal tensor that, similarly to that connecting in between $\vec{E}_{loc}$ and $\vec{E}_{i}$, it connects 
$\vec{P}$ with the macroscopic field $\vec{E}$,   
\begin{equation}
\label{matrix}
\begin{pmatrix}
P_x\\
P_y\\
P_z
\end{pmatrix}=\epsilon_0
\begin{pmatrix}
\chi_{xx} & 0 & 0\\
0 & \chi_{yy} & 0\\
0 & 0 & \chi_{zz}
\end{pmatrix}
\begin{pmatrix}
E_x\\
E_y\\
E_z
\end{pmatrix}
\end{equation}
In order to find the expressions for $\chi_{xx}$, $\chi_{yy}$, $\chi_{zz}$ and $E_x$ $E_y$, $E_z$, we will consider the case of $s$ and $p$ polarized incident waves.
\begin{figure}
\centerline{\includegraphics[width=9.cm]{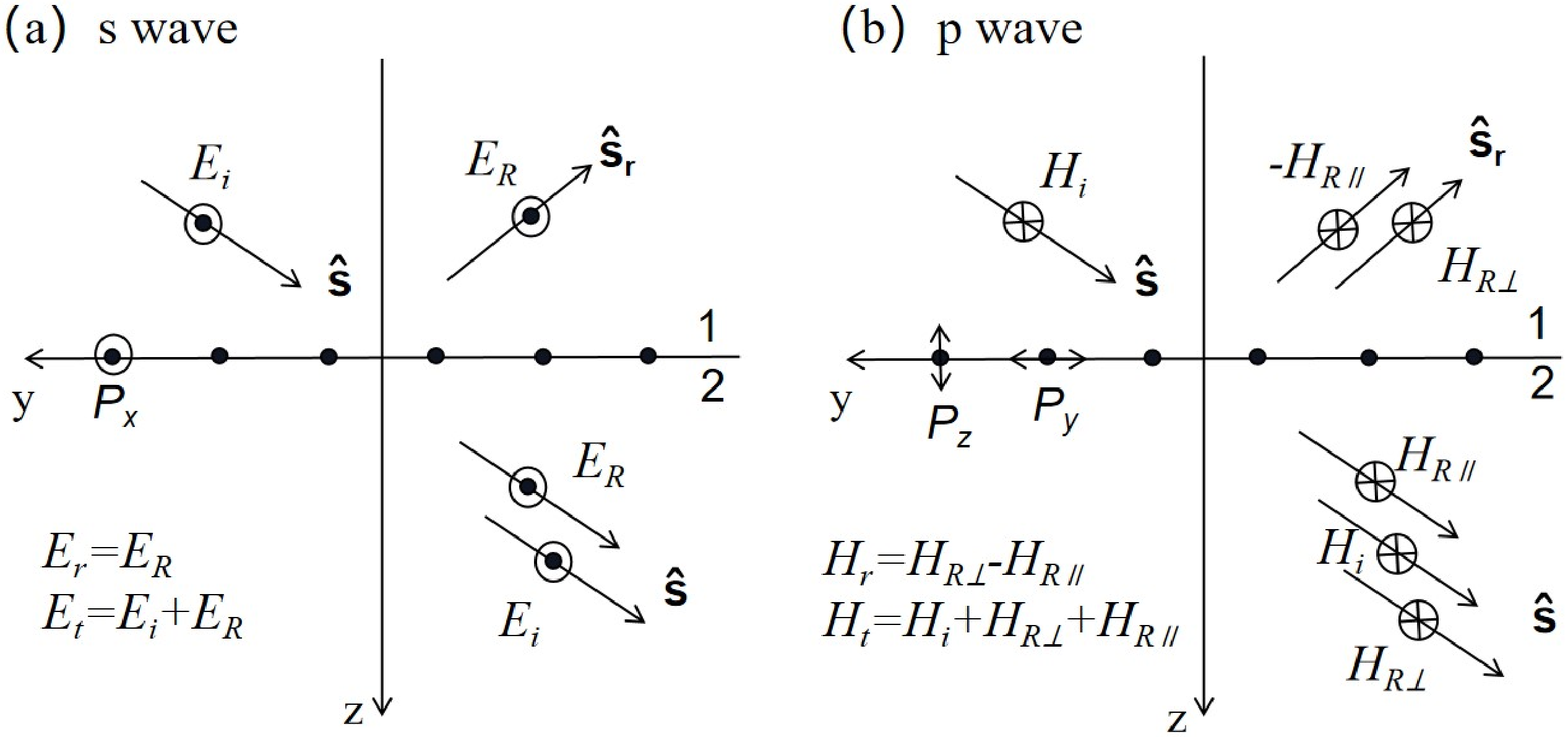}}
\caption{\label{Fig3} For an $s$ polarized wave, the total macroscopic electromagnetic field is the sum of the incident field plus the radiation reaction field due to the surface electric current associated to the surface polarization $P_x$. For a $p$ polarized wave, the total macroscopic electromagnetic field is the sum of the incident field plus the radiation reaction fields due to the surface electric current associated to $P_y$ and the surface magnetic current (see text) associated to $P_z$. The radiation reaction fields propagate along $\hat{s_r}$ (reflected field) and $\hat{s}$ (transmitted field).}
\end{figure}

\subsection{s-wave}

For an $s$ polarized incident wave the electric field oscillates along the $x$-direction (see Fig.\ref{Fig1}). We choose $\vec{E}_i=E_i \, \hat{x}$, $\vec{E}_r=E_r\, \hat{x}$ and $\vec{E}_t=E_t \, \hat{x}$ for the incident, reflected and transmitted beams.  Equations (\ref{Elocx})-(\ref{matrix}) imply that only the components $E_{loc, x}$, $P_x$ and $E_x$ are non-null in this case. This allows us to fix $\chi_{xx}$. The polarization vector varies in time since $\vec{E}_{loc} \propto e^{i\omega t}$, therefore, along the crystal plane, it gives rise to an in-plane surface electric current $J_{P_{x}}$, which is given by
\begin{equation}
 J_{P_{x}}=\partial P_x / \partial t=i\omega P_x=i\omega\epsilon_0 \alpha N E_{loc, x} 
\end{equation}
This surface current generates a macroscopic electromagnetic field that propagates in two different directions. The first goes along the unit vector $\hat{s}_r$ and it corresponds to the reflected field. The second one goes along the unit vector $\hat{s}$ and, in superposition with the incident field, it gives rise to the transmitted field (see Fig. \ref{Fig3}). 
The total macroscopic field is given by the incident field plus the macroscopic field generated by the dipoles. It must obey the following boundary conditions \cite{senior1987sheet,Idemen1990,kuester2003averaged,falkovsky2007optical,hanson2008dyadic,majerus2018electrodynamics}
\begin{align}
\label{boundary_x1}
&\hat{z} \wedge\left(\vec{E}_2-\vec{E}_1\right)=0 \\
\label{boundary_x2}
&\hat{z} \wedge\left(\vec{H}_2-\vec{H}_1\right)=J_{P_{x}}\, \hat{x} 
\end{align}
where the subscript 1 (2) refers to the limit of the macroscopic field when it approaches the crystal from above (below). 
These equations show that the component of the macroscopic electric field generated by the dipoles is continuous across the crystal surface whereas the component of the macroscopic magnetic field due to the dipoles is discontinuous. Using 
\begin{equation}
\eta \vec{H}=\hat{s} \wedge \vec{E}
\end{equation}
where $\eta$ is the wave impedance of the vacuum, and since $\hat{z}\wedge \hat{s} \wedge \vec{E}=-\vec{E}\cos\theta$, we can write Eqs. (\ref{boundary_x1}), (\ref{boundary_x2}) and (\ref{matrix}) in terms of the electric field components
\begin{align}
\label{component_x_boundary1}
E_{i}+E_{r} &= E_{t}  \\
\label{component_x_boundary2}
E_{i}-E_{r} &= E_{t}+\frac{\eta J_{P_{x}}}{\cos\theta}  \\
\label{component_x_boundary3}
E_{t} &= \frac{P_x}{\epsilon_0 \chi_{xx}}
\end{align}
We have three equations with three unknown variables: $E_{r}$, $E_{t}$, and $\chi_{xx}$. Defining the reflection and the transmission coefficients as $r_s=E_{r}/E_{i}$ and $t_s=E_{t}/E_{i}$, respectively, we obtain
\begin{align}
r_{s}  &= \frac{-i k \chi_{xx}}{i k \chi_{xx }+2 \cos \theta} \\
t_{s} &=1+r_{s} \\
\chi_{xx} &=\frac{4\pi a^3 N \alpha}{4\pi a^3-C_0 \alpha}  
\end{align}
The Fresnel coefficients for the $s$ polarized wave are the same already obtained in Ref. \cite{merano2016fresnel}. The surface susceptibility is the same obtained for normal incidence \cite{dell2016clausius,merano2017role}, showing that non-local effects are excluded in our theory.

Our microscopic theory allows us also to compute the radiation reaction electric field acting on the dipole at the origin of our reference frame \cite{merano2017role}, getting
\begin{align}
\label{s-current}
 E_{R_{x}} & = \frac{i k a C_{1} \alpha }{4\pi a^{3}\cos \theta }\,E_{loc,x } = -\frac{\eta J_{P_{x}}}{2\cos \theta}  \end{align}
 This is clearly a macroscopic quantity as well as $J_{P_{x}}$. We now show that it corresponds to the macroscopic field generated by the oscillating dipoles. Using Eq. (\ref{s-current}) in Eq. (\ref{component_x_boundary2}) we can rewrite Eqs. (\ref{component_x_boundary1})-(\ref{component_x_boundary3}) getting 
\begin{align}
\label{solutions_x_boundary}
E_{r} & = E_{R_{x}}  \nonumber \\
E_{t} & = E_{i}+E_{R_{x}} \nonumber \\
\chi_{xx} & = \frac{P_x}{\epsilon_0 (E_{i}+E_{R_{x}})}
\end{align}
finding that $E_{R_{x}}$ is equal to $E_{r}$ and, in superposition with $E_{i}$, it gives the transmitted field. For that reason we can identify $E_{R_{x}}$ with the macroscopic electric field generated by the oscillating dipoles that propagates along $\hat{s}$ and $\hat{s}_r$ directions (see Fig. \ref{Fig3}), while, from the last equation, the full macroscopic electric field on the crystal plane can be identified with the sum of the radiation reaction field and the incident field, namely $E_{x} =E_{i}+E_{R_{x}}$.

\subsection{p-wave}

For a $p$ polarized wave the incident, reflected and transmitted magnetic fields oscillate along $x$-direction (see Fig.\ref{Fig1}). We choose, therefore, $\vec{H}_i=-H_i\, \hat{x}$, $\vec{H}_r=-H_r \, \hat{x}$ and $\vec{H}_t=-H_t \, \hat{x}$. Since we computed the local electric field, it is useful to first write explicitly the incident, reflected and transmitted electric fields 
\begin{align}
\label{fields}
\vec{E}_i&=E_i\cos{\theta}\, \hat{y} +E_i\sin{\theta} \, \hat{z}=E_{iy} \, \hat{y} +E_{iz} \, \hat{z} \nonumber \\
\vec{E}_r&= -E_r\cos{\theta}\, \hat{y} +E_r\sin{\theta} \, \hat{z}=E_{ry} \, \hat{y} +E_{rz} \, \hat{z} \nonumber \\
\vec{E}_t&=E_t\cos{\theta}\, \hat{y} +E_t\sin{\theta} \, \hat{z}= E_{ty} \, \hat{y} +E_{tz} \, \hat{z}
\end{align}
where $E_{i}=\eta H_i$, $E_{r}=\eta H_r$, $E_{t}=\eta H_t$. Equations (\ref{Elocx})-(\ref{matrix}) imply that we can fix separately $\chi_{y y}$ and $\chi_{z z}$. In this case we have one varying macroscopic surface polarization along the crystal plane ($P_y$) and one perpendicular to the crystal plane ($P_z$). These two components induce two surface currents $J_{P_{y}}$ and $J_{P_{z}}$ respectively. They generate two macroscopic fields propagating along the $\hat{s}$ direction and two along the $\hat{s_r}$ direction. According to the superposition principle the total macroscopic field is the sum of the incident field and the macroscopic fields generated by the two currents considered separately (see Fig. \ref{Fig3}). The expression of the in-plane electric current is
\begin{equation}
 J_{P_{y}}=\partial P_y / \partial t=i\omega P_y=i\omega\epsilon_0 \alpha N E_{loc,y} 
\end{equation}
In the radiation zone the electromagnetic field, due to an electric dipole oscillating in the $ \hat{z} $ direction, is identical to an electromagnetic field due to a magnetic dipole oscillating in the $ -\hat{x} $ direction \cite{jackson1999classical}. Hence, a polarization $P_z$ generates an out-of-plane electric surface current equivalent to an in-plane magnetic surface current, responsible for the discontinuity of the macroscopic electric field \cite{senior1987sheet,Idemen1990,kuester2003averaged,majerus2018electrodynamics}. Using $P_z\propto E_{loc,z}\propto E_{i,z}$ and $\vec{E}_i\propto e^{i(\omega t+ky\sin\theta)}$, we have
\begin{equation}
J_{P_{z}}\, \hat{x}= -\frac{1}{\epsilon_{0}} \hat{z} \wedge \vec{\nabla}P_z 
=\frac{i k\, \sin{\theta}}{\epsilon_0} {P_{z}}\, \hat{x}
\end{equation} 
We must solve two separate sets of boundary conditions for the macroscopic field 
\begin{equation}
\label{boundary_y}
\renewcommand{\arraystretch}{2}
\setlength{\arraycolsep}{0pt}
\begin{array}{
  >{\displaystyle}l   
  >{\displaystyle{}}l 
  @{\hspace{1em}}     
  |                   
  @{\hspace{1em}}     
  >{\displaystyle}l   
}
&\hat{z} \wedge\left(\vec{E}_2-\vec{E}_1\right)=0  & \hat{z} \wedge\left(\vec{E}_2-\vec{E}_1\right)=J_{P_{z}}\, \hat{x} \\ \vspace{1em}
&\hat{z} \wedge\left(\vec{H}_2-\vec{H}_1\right)=J_{P_{y}}\, \hat{y} & \hat{z} \wedge\left(\vec{H}_2-\vec{H}_1\right)=0 \\
\end{array}
\end{equation}
In terms of the magnetic field components, Eqs. (\ref{boundary_y}) and (\ref{matrix}) become
\begin{equation}
\label{boundary_yH}
\renewcommand{\arraystretch}{1.5}
\setlength{\arraycolsep}{0pt}
\begin{array}{
  >{\displaystyle}r   
  >{\displaystyle{}}l 
  @{\hspace{0.1em}}     
  |                   
  @{\hspace{0.2em}}     
  >{\displaystyle}l   
}
H_{i}-H_{r\Vert} &= H_{t\Vert}   & H_{i}-H_{r\bot} =H_{t\bot}+\frac{J_{P_z}}{\eta \cos{\theta}}\\
H_{i}+H_{r\Vert} &= H_{t\Vert}+J_{P_y} & H_{i}+H_{r\bot} = H_{t\bot} \\
\eta \cos{\theta} \  H_{t\Vert}  &= \frac{P_y}{\epsilon_0 \chi_{yy}} & \eta \sin{\theta} \  H_{t\bot}  = \frac{P_z}{\epsilon_0 \chi_{zz}}\\
\end{array}
\end{equation}
where $H_{r\Vert}$ ($H_{r\bot}$) is the contributions to the reflected field due to the in-plane (out-of-plane) surface current and $H_{t\Vert}$ and $H_{t\bot}$ are the total macroscopic magnetic fields immediately below the crystal for the two cases considered separately. We have,
then, the following solutions
\begin{equation}
\renewcommand{\arraystretch}{2.5}
\setlength{\arraycolsep}{0pt}
\begin{array}{
  >{\displaystyle}r   
  >{\displaystyle{}}l 
  @{\hspace{1em}}     
  |                   
  @{\hspace{0.3em}}     
  >{\displaystyle}l   
}
&\chi_{yy} =\frac{4\pi a^3 N \alpha}{4\pi a^3-C_0 \alpha} & \chi_{zz} =\frac{2\pi a^3 N \alpha}{2\pi a^3+C_0 \alpha} \\
&\frac{H_{r\Vert}}{H_{i}}= \frac{i k \chi_{yy} \cos{\theta}}{i k \chi_{yy }\cos{\theta}+2} & \frac{H_{r\bot}}{H_{i}} = \frac{-i k \chi_{zz} \sin{\theta}\tan{\theta}}{i k \chi_{zz }\sin{\theta}\tan{\theta}+2} \\
&\frac{H_{t\Vert}}{H_{i}} =\frac{2}{i k \chi_{yy }\cos{\theta}+2}&\frac{H_{t\bot}}{H_{i}}=\frac{2}{i k \chi_{zz }\sin{\theta}\tan{\theta}+2}\\ 
\end{array}
\end{equation}
This finding is a very interesting result because it shows that our crystal has an isotropic in-plane surface susceptibility ($\chi _{\Vert}= \chi_{xx}= \chi_{yy} $) different from the out-of-plane susceptibility ($\chi _{\bot}= \chi_{zz}$), namely we are dealing with an uniaxial crystal. We can also evaluate the ratio $\chi_\bot/ \chi_\Vert$. Choosing the value of $a = 0.26$ nm, for atom polarizabilities $\alpha$ typically varying in the range between 1 and 30 $\rm cm^{-3}$, the $\chi_\bot/ \chi_\Vert$ ratio varies in between 0.93 and 0.1 (0.94 and 0.18) for a triangular lattice (for a square lattice). This values are of the same order of magnitude of what has been observed \cite{xu2021optical}.

The full reflected field from the crystal is $H_{r}=H_{r\Vert}+H_{r\bot}$ (see Fig. \ref{Fig3}), so that we can finally derive the reflection coefficient for 2D crystals, $r_p={H_{r}}/{H_{i}}$, getting
\begin{equation}
r_p=\frac{i k \chi_{yy} \cos{\theta}}{i k \chi_{yy }\cos{\theta}+2}-\frac{i k \chi_{zz} \sin{\theta}\tan{\theta}}{i k \chi_{zz }\sin{\theta}\tan{\theta}+2}
\end{equation}
We compute now the radiation-reaction electric field acting on the dipole at the origin along the $y$ and $z$ directions
\begin{align}
E_{R_{y}}& = \frac{i k a C_{1} \alpha \cos \theta}{4 \pi a^{3}} E_{loc, y} = -\frac{\eta J_{y} \cos \theta}{2} 
\end{align}
and
\begin{align}
E_{R_{z}} & = \frac{i k a C_{1} \alpha (\sin \theta)^2}{4 \pi a^{3}\cos{\theta}} E_{l o c_{z}}  = -\frac{i k (\sin \theta)^2 }{2 \epsilon_{0}\cos{\theta}} P_{z}\nonumber \\ & = -\frac{\tan \theta}{2 \epsilon_{0}} \frac{\partial}{\partial y} P_{z}= -\frac{\tan \theta}{2}J_{P_{z}}.
\end{align}
and we show that these macroscopic quantities correspond to the macroscopic fields generated by the oscillating dipoles. We can rewrite Eqs. (\ref{boundary_yH}) in terms of the electric fields
\begin{equation}
\renewcommand{\arraystretch}{1.5}
\setlength{\arraycolsep}{0pt}
\begin{array}{
  >{\displaystyle}r   
  >{\displaystyle{}}l 
  @{\hspace{1em}}     
  |                   
  @{\hspace{1em}}     
  >{\displaystyle}l   
}
&E_{iy}+E_{ry\Vert} = E_{ty\Vert} & E_{iz}+E_{rz\bot} = E_{tz\bot} \\
&E_{iy}-E_{ry\Vert} =E_{ty\Vert} -2E_{R_y} & E_{iz}-E_{rz\bot} =E_{tz\bot}-2 E_{R_z} \\
&\epsilon_0  \chi_{yy} E_{ty\Vert} = P_y & \epsilon_0  \chi_{zz} E_{ty\bot} = {P_z}\\ 
\end{array}
\end{equation}
which give the following solutions
\begin{equation}
\label{solutions_yz_boundary}
\renewcommand{\arraystretch}{1.5}
\setlength{\arraycolsep}{0pt}
\begin{array}{
  >{\displaystyle}r   
  >{\displaystyle{}}l 
  @{\hspace{1em}}     
  |                   
  @{\hspace{1em}}     
  >{\displaystyle}l   
}
&E_{ry\Vert} = E_{R_y} & E_{rz\bot} = E_{R_z} \\
&E_{ty\Vert} =E_{iy}+E_{R_y} & E_{tz\bot} =E_{iz}+E_{R_z} \\
&\chi_{yy} = \frac{P_y}{\epsilon_0 (E_{iy}+E_{R_y})} & \chi_{zz} = \frac{P_z}{\epsilon_0 (E_{iz}+E_{R_z})}\\ 
\end{array}
\end{equation}
We can interpret these equations in the following way. The in-plane (out-of-plane) electric surface current $J_{P_{y}}$ ($J_{P_{z}}$) generates a macroscopic electromagnetic field propagating along the $\hat{s}$ and $\hat{s}_r$ directions. We identify $E_{R_y}$ ($E_{R_z}$) as the electric field component along the $y$-direction ($z$-direction) of this macroscopic field. From Eqs. (\ref{fields})), we can write the first two equations on the left-hand-side of Eqs. (\ref{solutions_yz_boundary}) as
\begin{align}
\label{magnetic y}
H_{r\Vert} &=-\frac{E_{R_y}}{\eta \cos{\theta}}=-H_{R\Vert}  \nonumber \\
H_{t\Vert} &= H_i+\frac{E_{R_y}}{\eta \cos{\theta}}= H_i+H_{R\Vert} 
\end{align}
and the first two equations on the right-hand-side of Eqs. (\ref{solutions_yz_boundary}) as
\begin{align}
\label{magnetic z}
H_{r\bot} &=\frac{E_{R_z}}{\eta \sin{\theta}}=H_{R\bot}  \nonumber \\
H_{t\bot} &= H_i+\frac{E_{R_z}}{\eta \sin{\theta}}= H_i+H_{R\bot} 
\end{align}
The full reflected field (see Fig. \ref{Fig3}) is, therefore,
\begin{equation}
H_r=-H_{R\Vert}+H_{R\bot}
\end{equation} 
while the full transmitted field is given by the superposition of the incident field and the two macroscopic fields propagating in the $\hat{s}$ direction (see Fig. \ref{Fig3}), generated by the two currents,  
\begin{equation}
H_t=H_i+H_{R\Vert}+H_{R\bot}
\end{equation} 
We can finally derive the transmission coefficient, defined by $t_p=H_t/H_i$, getting
\begin{equation}
    t_p=1-\frac{i k \chi_{yy} \cos{\theta}}{i k \chi_{yy }\cos{\theta}+2}-\frac{i k \chi_{zz} \sin{\theta}\tan{\theta}}{i k \chi_{zz }\sin{\theta}\tan{\theta}+2}
\end{equation}
Finally the equations in the last line of Eqs. (\ref{solutions_yz_boundary}) provide the macroscopic electric fields $E_y$ and $E_z$. They are related to the radiation-reaction electric fields. Along the $y$-direction we have $E_y=E_i \cos{\theta}+E_{R_y}$ while along the $z$-direction we have $E_z=E_i \sin{\theta}+E_{R_z}$.

\section{CONCLUSION}

We have shown that a 2D crystal, microscopically composed by atoms with an isotropic polarizability $\alpha$, has both a macroscopic $\chi_\Vert$ and a macroscopic $\chi_\bot$ with $\chi_\bot \neq \chi_\Vert$ and $\chi_\bot \neq 0$, in contrast to some theoretical models, 
generally used to derive reflection and transmission coefficients for these materials, 
which assume either a null $\chi_\bot$ \cite{hanson2008dyadic,falkovsky2007optical, Zhan2013, merano2016fresnel} or isotropy \cite{blake2007making, li2014measurement, Liu14, Li14, Morozov15}. Moreover we have confirmed that $\chi_\bot<\chi_\Vert$ and that the ratio $\chi_\bot / \chi_\Vert$ is compatible with experimental observations.

Our microscopic theory provides both the local and the radiation-reaction fields when an incoming electromagnetic plane wave shines on a 2D material. The radiation-reaction fields coincide with the macroscopic fields scattered by the crystal, and they contribute to the reflected and the transmitted radiations. This analysis allows us, then, to derive the complete set of Fresnel coefficients and their dependence on the angle of incidence.  

In particular we describe the reflection of an incident $p$ wave, as obtained by the superposition of the scattered electromagnetic fields due to an in-plane electric and an in-plane magnetic surface currents. The first current is related to a macroscopic surface polarization oscillating in the crystal plane, the second one to a macroscopic surface polarization oscillating perpendicularly to the crystal plane. Combining the boundary conditions for these two currents we provide a simple expression for the macroscopic electric fields along the $y$ and $z$ directions, which turn to be the sum of the related components of the incident field with the radiation-reaction field along the same directions. 
For an $s$ polarized wave, only an in-plane electric surface current is involved, so the reflected field is given by the scattered electromagnetic field due to only an in-plane current and the macroscopic electric field is the sum of the incident field plus the radiation reaction field.

\begin{acknowledgments}
Y.H acknowledges financial support from China Scholarship Council. L.D was in charge of the microscopic theory, M.M of the macroscopic theory, all the authors discussed the physical model and wrote the paper. 
\end{acknowledgments}

\nocite{*}

\bibliography{articolo1}

\end{document}